\documentclass[12pt,a4paper]{article}
\usepackage{amsmath}
\usepackage{amsfonts}
\usepackage{amssymb}
\usepackage{textcomp}
\usepackage[dvips]{graphicx}
\usepackage{epsfig}
\usepackage{bm}
\usepackage{dcolumn}
\usepackage{amsfonts}
\usepackage{color}
\usepackage[left=2cm,right=2cm,top=2cm,bottom=2cm]{geometry}
\begin{document}
\title{\textbf{Generalization of cosmological attractor approach to the Einstein-Gauss-Bonnet gravity}}
\author{E.O.~Pozdeeva$^{a,}$\footnote{E-mail: pozdeeva@www-hep.sinp.msu.ru} \\
\small $^a$ Skobeltsyn Institute of Nuclear Physics, Lomonosov Moscow State University,\\ \small  Leninskie Gory~1, 119991, Moscow, Russia}

\date{ \ }
\maketitle
\begin{abstract}
We construct models with the Gauss-Bonnet term multiplied to a function of the scalar field leading to an inflationary scenario. The consideration is related to the slow-roll approximation. The cosmological
attractor approach gives the spectral index of scalar perturbations which is in a good agreement
with modern observation and allows variability for tensor-to-scalar ratio. We reconstruct models with variability of parameters which allow to reproduce cosmological attractor predictions for inflationary parameters
in the leading order of $1/N$ approximation in the Einstein-Gauss-Bonnet gravity.
\end{abstract}

\section{Introduction}

The solution to problems of horizon, smoothness, flatness and monopole which are
related with the hot big-bang model was proposed by the introduction of inflation \cite{problems}.
The $R^2$  inflationary predictions \cite{predictions} in the leading approximation
in terms of inverse e-folding numbers $1/N$ for spectral index $n_s$ and tensor-to-scalar
ratio $r$:
\begin{equation}n_s\simeq1-\frac{2}{N},\quad r\simeq\frac{12}{N^2}\label{predictionR2}\end{equation}
are in good agreement with Planck 2018 data\footnote{There exist two variant for interpretation of relation between time derivative
and e-folding number derivative: 1) $\displaystyle\frac{d}{dt}=H\displaystyle\frac{d}{dN_e}$ and 2) $\displaystyle\frac{d}{dt}=-H\displaystyle\frac{d}{dN}$.\\
In the case  of the first type formulation,  the  inflation interval in the e-folding formulation is $-65<{N_e}<0$. \\
In the case of the second type formulation, inflation interval in e-folding formulation  $0<{N}<65$.
The second formation was applied in cosmological attractor approximation  \cite{Kallosh:2013hoa} and we follow to the second formulation with $N=-\ln(a)$.} \cite{Akrami:2018odb}.

The inflationary scenario motivated by the Standard Model of particles physics, the Higgs-driven inflation \cite{BezrukovShaposhnikov}, leads to the same prediction. The  Higgs-driven inflation belongs to  class of cosmological attractors \cite{Kallosh:2013hoa}, which generalize the prediction \eqref{predictionR2}.

The cosmological attractor models predict the same values of observable parameters $n_\mathrm{s}$ and $r$ in the leading $1/N$ approximation:
\begin{equation}n_s\simeq1-\frac{2}{N+N_0},\quad r\simeq\frac{12C_\alpha}{(N+N_0)^2}\label{prediction}\end{equation}
where $C_\alpha$ and $N_0\ll 60$ are constants.

The Higgs-driven inflation was generalized to multi-field inflationary scenarios \cite{Dubinin:2017irg}, for which the cosmological attractor approximation is appropriate \cite{Dubinin:2017irqA}.

At present, the interest in inflationary scenarios in the cosmological models with Gauss-Bonnet term is growing
\cite{Fomin:2020hfh,vandeBruck:2017voa,Guo:2010jr}. In the present paper, we construct a gravity
model with the Gauss-Bonnet term multiplied to a function of a scalar field which allows to reconstruct expressions for spectral
index and tensor-to-scalar ratio from cosmological attractor models in the slow-roll regime. This model includes several
constants with variable values. Therefore, we construct a family of models with different values of the constants.
We consider the scalar power spectral amplitude and estimate possible values of model parameters
 using modern observational data \cite{Akrami:2018odb}

The paper is organized as follows. In section 2, we  reformulate the problem of the slow-roll regime in Einstein-Gauss-Bonnet gravity
 in terms of e-folding numbers.
In section 3, we apply our reformulation to construct model with variable values of parameters which leads to the cosmological attractor approximation for inflationary parameters. To satisfy observational data we introduce restriction to model parameters. In Conclusions we summaries our results.

\section{Slow-roll regime in Einstein-Gauss-Bonnet gravity}
We consider the  model with the Gauss-Bonnet term multiplied to a function of the scalar field $\phi$:
\begin{equation}
\label{action1}
S=\int d^4x\sqrt{-g}\left[\frac{R}{2}-\frac{g^{\mu\nu}}{2}\partial_\mu\phi\partial_\nu\phi-V(\phi)-\frac{\xi(\phi)}{2}{\cal G}\right],
\end{equation}
where ${\cal G}=R_{\mu\nu\rho\sigma}R^{\mu\nu\rho\sigma}-4R_{\mu\nu}R^{\mu\nu}+R^2$.
The model is presented in the Planckian unit: $h=c=8\pi G=1$. Application of the variation principe leads to the following system of equations \cite{Guo:2010jr} in spatially flat FLRW metric with $ds^2=-dt^2+a^2(t)(dx^2+dy^2+dz^2)$:
\begin{eqnarray}
               &&6H^2 = \dot{\phi}^2+2V+24\dot{\xi}H^3, \\
                &&2\dot{H} ={}-\dot{\phi}^2+4\ddot{\xi}H^2+4\dot{\xi}H\left(2\dot{H}-H^2\right), \\
                &&\ddot{\phi}+3H\dot{\phi}+V_{,\phi}+12\xi_{,\phi}H^2\left(\dot{H}+H^2\right)=0,
\end{eqnarray}
where $H=\dot{a}/a$, the dot means the derivative of time: $\dot{A}=dA/dt$ .
 We consider the model \eqref{action1} in FLRW metric in the slow-roll regime \cite{Guo:2010jr}:
 \begin{equation}
 \dot{\phi}^2\ll V, \quad |\ddot{\phi}|\ll 3H|\dot{\phi}|,\quad 4|\dot{\xi}|H\ll1,\quad |\ddot{\xi}|\ll|\dot{\xi}|H\label{slow-roll},
 \end{equation}
in which  of the equations of motion are:
 \begin{eqnarray}
               &&H^2\simeq\frac{V}{3}, \quad\dot{H} \simeq-\frac{\dot{\phi}^2}{2}-2\dot{\xi}H^3,\quad\dot{\phi}\simeq-\frac{V_{,\phi}+12\xi_{,\phi}H^4}{3H}\label{dot(phi)}.
\end{eqnarray}
The  slow-roll parameters are:
\begin{eqnarray}
  \epsilon_1 &=&-\frac{\dot{H}}{H^2},\quad \epsilon_{i+1}= \frac{d\ln|\epsilon_i|}{d\ln a},\quad i\geqslant 1, \\
  \delta_1&=& 4\dot{\xi}H,\quad \delta_{i+1}=\frac{d\ln|\delta_i|}{d\ln a},\quad i\geqslant 1.
\end{eqnarray}
To get cosmological attractor generalization we consider the model in slow-roll regime using the e-folding number representation and designation $\frac{dA}{dN}=A^\prime$:
\begin{eqnarray}
  &&(\phi^\prime)^2\simeq\frac{V^\prime}{V}+\frac43\xi^\prime V=\frac{\left(H^2\right)^\prime}{H^2}+4H^2\xi^\prime. \label{phiPrime}
\end{eqnarray}
We present the slow-roll parameters  in terms of $H^2$ and $\xi$:
\begin{eqnarray}
  &&\epsilon_1=\frac12\frac{(H^2)^\prime}{H^2}, \\
  &&\epsilon_2=\frac{(H^2)^\prime}{H^2}-\frac{(H^2)^{\prime\prime}}{(H^2)^\prime}=2\epsilon_1-\frac{(H^2)^{\prime\prime}}{(H^2)^\prime},\\
  &&\delta_1=-4H^2\xi^\prime, \\
  && \delta_2=-\frac{(H^2)^\prime}{H^2}-\frac{\xi^{\prime\prime}}{\xi^\prime}=-2\epsilon_1-\frac{\xi^{\prime\prime}}{\xi^\prime}.
\end{eqnarray}
The slow-roll approximation requires $|\epsilon_i|\ll1,$ $|\delta_i|\ll1.$

The questing about restrictions to inflation scenarios related with speed of sound in Einstein-Gauss-Bonnet gravity  was considered in \cite{Hikmawan:2015rze}. There is a wonderful properties of  slow-roll regime: speed of sound is real.
Accordingly with  \cite{Odintsov:2020zkl} the  speed of sound square can be presented in the form:
\begin{equation}\label{SOUND}
    c_A^2=1+\Delta c_A^2,\quad\mbox{where}\quad \Delta c_A^2=-\frac{2\delta_1^2\epsilon_1}{3\delta_1^2+2(2\epsilon_1-\delta_1)(1+\delta_1)}.
 \end{equation}
In general case of slow-roll regime $\Delta c_A^2\simeq-(\delta_1^2\epsilon_1)/(2\epsilon_1-\delta_1)\ll1$.
If $2\epsilon_1\approx\delta_1$, then $\Delta c_A^2\simeq-2\epsilon_1/3\ll 1$. Thus, we can conclude $ c_A^2>0$ in slow-roll regime.

In \cite{Guo:2010jr} the spectral index of scalar perturbations and the tensor-to-scalar ratio are presented
in terms of slow-roll parameters:
\begin{eqnarray}
  &&n_s=1-2\epsilon_1-\frac{2\epsilon_1\epsilon_2-\delta_1\delta_2}{2\epsilon_1-\delta_1},\label{n_s_i} \\
  &&r=8|2\epsilon_1-\delta_1|\label{r_i},
\end{eqnarray}
the expression for amplitude \cite{vandeBruck:2017voa} in terms of inflationary parameters is  as follows:
\begin{equation}
A_s\simeq\frac{2H^2}{\pi^2 r}.\label{As}
\end{equation}

We simplify the expression for  spectral index of scalar perturbations  \eqref{n_s_i} remembering that $\epsilon_2=-\epsilon_1^\prime/\epsilon_1$, $\delta_2=-\delta_1^\prime/\delta_1$  and tensor-to-scalar ratio \eqref{r_i} using \eqref{phiPrime}. So, the  inflationary parameters can be presented in the  following form:
\begin{eqnarray}
  &&n_s=1-2\epsilon_1+\frac{r^\prime}{r}  \label{n_r(phi)}\,,\\
  &&r=8|2\epsilon_1-\delta_1|=8\left(\frac{(H^2)^\prime}{H^2}+4H^2\xi^\prime\right)=8(\phi^\prime)^2\label{r(phi)}\,.
\end{eqnarray}

 \section{Generalization of the cosmological attractor method}\label{alpha}
Accordingly to inflationary parameters of cosmological--attractor models without the Gauss-Bonnet term~\eqref{prediction} spectral index includes only logarithmic derivative of tensor-to-scalar ratio
\begin{equation}
\frac{r^\prime}{r}=-\frac{2}{N+N_0}, \quad\mbox{and}\quad n_s\approx1+\frac{r^\prime}{r}.
\end{equation}
in the leading order of  $1/N$  approximation.  The  model without the Gauss-Bonnet term and exponential potential leading to \eqref{prediction} was considered in \cite{Mukhanov:2013tua}. In next subsection we generalize this model to the Einstein-Gauss-Bonnet gravity.
\subsection{Exponential form}
To generalize cosmological attractor approximation to inflationary models with the Gauss-Bonnet term we compare  \eqref{r(phi)} with \eqref{prediction}:
\begin{equation}
\frac{r}{8}=\frac{(H^2)^\prime}{H^2}+4H^2\xi^\prime=\frac{3C_\alpha}{2(N+N_0)^2}.\label{rG}
\end{equation}
For simplicity we suppose that all terms in this equation are proportional to $1/(N+N_0)^2$ and get the same approximation of slow-roll parameter $\epsilon_1$ in leading $1/N$ order:
\begin{eqnarray}
H^2=H_0^2\exp\left(-\frac{3C_\beta}{2(N+N_0)}\right)\label{H2(N)},\quad  \xi=\xi_0\exp\left(\frac{3C_\beta}{2(N+N_0)}\right)\label{xi(N)},
\end{eqnarray}
where $C_\beta$ is a constant.
We  substitute  \eqref{xi(N)}  to  \eqref{rG} and get:
\begin{equation}
\frac{r}{8}=\frac{3C_\beta}{2(N+N_0)^2}\left(1-4\xi_0H_0^2\right), \end{equation}
fixing a relation between $C_\alpha$ and $C_\beta$:
\begin{equation}
C_\beta=\frac{C_\alpha}{1-4\xi_0H_0^2},\quad H_0^2\neq\frac{1}{4\xi_0}.\label{C}\end{equation}
Accordingly   \eqref{r(phi)} the derivative of field is related with e-folding number:
\begin{equation}
(\phi^\prime)^2=\frac{3C_\alpha}{2(N+N_0)^2}\quad \mbox{or }\quad \phi^\prime=\omega_\phi\sqrt{\frac{3C_\alpha}{2}}(N+N_0)^{-1},\quad \omega_\phi=\pm1
\end{equation}
from here
\begin{equation}
\phi=\omega_\phi\sqrt{\frac{3C_\alpha}{2}}\ln\left(\frac{N+N_0}{N_\phi}\right)\quad\mbox{or}\quad N+N_0=N_\phi \exp\left(\omega_\phi\sqrt{\displaystyle\frac{2}{3C_\alpha}}\phi\right)\,.
\label{phi(N)}
\end{equation}
Using \eqref{dot(phi)}, \eqref{H2(N)} and  \eqref{phi(N)} we construct family of the models with the Gauss-Bonnet interaction and potential with variable parameter $C_\alpha$:
\begin{equation}
V=3H_0^2\exp\left(-\frac{3}{2}\frac{C_\beta}{N_\phi}\exp\left(-\omega_\phi\sqrt{\frac{2}{3C_\alpha}}\phi\right)\right),\quad
\xi=\xi_0\exp\left(\frac{3}{2}\frac{C_\beta}{N_\phi}\exp\left(-\omega_\phi\sqrt{\frac{2}{3C_\alpha}}\phi\right)\right)\label{answear}
\end{equation}
leading to appropriate inflationary scenarios. This model is generalization of the general relativity model obtained in \cite{Mukhanov:2013tua}.

We would like to compare inflationary parameters of obtained model \eqref{answear} with inflationary parameters of the following model:%
\begin{equation}
V=3H_0^2\left(1-\frac{3C_\beta}{4N_\phi}\exp\left(-\omega_\phi\sqrt{\frac{2}{3C_\alpha}}\phi\right)\right)^2\quad \xi=\xi_0\left(1-\frac{3C_\beta}{4N_\phi}\exp\left(-\omega_\phi\sqrt{\frac{2}{3C_\alpha}}\phi\right)\right)^{-2}.\label{large fields answear}
\end{equation}
In this model the relation between e-folding numbers and  fields values is differ from \eqref{phi(N)} and can be presented in the form:
\begin{eqnarray}
  && \frac{N+N_0}{N_\phi}=\exp\left(\omega_\phi\sqrt{\frac{2}{3C_\alpha}}\phi\right)-\frac{3}{4}\frac{C_\beta}{N_\phi}\omega_\phi\sqrt{\frac{2}{3C_\alpha}}\phi \\
  &&\phi=-\omega_\phi\sqrt{\frac{3C_\alpha}{2}}\left({\it LambertW}\left(-\frac{4N_\phi}{3C_\beta}\exp\left(-\frac{4N}{3C_\beta}\right)\right)+\frac{4N}{3C_\beta}\right).\label{N(phi)approximated model}
\end{eqnarray}
Here we should note that if $\omega_\phi=+1$, then   $\exp\left(\omega_\phi\sqrt{\frac{2}{3C_\alpha}}\phi\right)-\frac{3}{4}\frac{C_\beta}{N_\phi}\omega_\phi\sqrt{\frac{2}{3C_\alpha}}\phi\simeq\exp\left(\omega_\phi\sqrt{\frac{2}{3C_\alpha}}\phi\right)$
in the large fields expansion and relation \eqref{N(phi)approximated model} can be roughly approximated to \eqref{phi(N)}.

\subsection{Inflationary parameters}
In this subsection, we get expressions for inflationary parameters in terms of  fields.
The tensor-to-scalar ratio  and spectral index of scalar perturbations can be presented in the following form \cite{Guo:2010jr}:
\begin{equation}
r=8Q^2,\quad n_s=1-Q\frac{V_{\phi}}{V}+2Q_{,\phi},\label{p(phi)}
\end{equation}
where $Q=\frac{V_{,\phi}}{V}+\frac{4\xi_{,\phi}V}{3}$.
 We consider \eqref{answear} and \eqref{large fields answear} to check correspondence of
expression for inflationary parameters. In the comparing analysis    we suppose
$N_\phi={3C_\beta}/{4}$ and $\omega_\phi=1$ for simplicity and consider the models:
\begin{eqnarray}
&&1.\quad V=3H_0^2\exp\left(-2\exp\left(-\sqrt{\frac{2}{3C_\alpha}}\phi\right)\right),\quad\xi=\xi_0\exp\left(2\exp\left(-\sqrt{\frac{2}{3C_\alpha}}\phi\right)\right),\\\label{for analis}
&&2.\quad V=3H_0^2\left(1-\exp\left(-\sqrt{\frac{2}{3C_\alpha}}\phi\right)\right)^2,\quad\xi=\xi_0\left(1-\exp\left(-\sqrt{\frac{2}{3C_\alpha}}\phi\right)\right)^{-2}.\label{large fields answear fix constant}
\end{eqnarray}
We use \eqref{p(phi)} to calculate the inflationary parameters for the model \eqref{for analis}
\begin{eqnarray}
1.\quad  &&n_s=1+\frac{8\left( 4{{H_0}}^{2}{\xi_0}-1 \right)}{3C_\alpha}\,\, {\exp\left(-\sqrt{\frac{2}{3C_\alpha}}\phi\right)}\left(1+{\exp\left(-\sqrt{\frac{2}{3C_\alpha}}\phi
\right)} \right),\\
  &&r=\frac{64\left( 4\,{{H_0}}^{2}{\xi_0}-1 \right)^{2}}{3C_\alpha}{\exp}\left(-2\sqrt
{\frac{2}{3C_\alpha}}\phi\right)
\end{eqnarray}
and for the model  \eqref{large fields answear fix constant}
\begin{eqnarray}
2.\quad\quad\quad  &&\tilde{n}_s=1+{\frac {8\left( 4{{H_0}}^{2}{\xi_0}-1 \right) {\exp\left(-\sqrt{\frac{2}{3C_\alpha}}\phi\right)}\left(1+{\exp\left(-\sqrt {\frac{2}{3C_\alpha}}\phi\right)}\right)}{3C_\alpha\left( 1-\exp\left(-\sqrt {\frac{2}{3C_\alpha}}\phi\right)\right) ^{2}}}, \\
  &&\tilde{r}={\frac {64\left( 4\,{{H_0}}^{2}{\xi_0}-1\right) ^{2}{\exp}\left(-2\sqrt{\frac{2}{3C_\alpha}}\phi\right)}{ 3C_\alpha\left(1-\exp\left(-\sqrt{\frac{2}{3C_\alpha}}\phi\right)\right)^{2}}}.
\end{eqnarray}

In the case of large field $\phi$ the expressions for $r$ and $\tilde{r}$ are coincide up to second order, the expressions for $n_s$ and $\tilde{n}_s$ are coincide up to first order of $\exp\left(-\sqrt{\frac{2}{3C_\alpha}}\phi\right)$.  The precision coincides with sensibility of cosmological attractor approximation \eqref{prediction}. To satisfy the proposal sensibility we can write $$n_s\simeq1+\frac{8\left(4H_0^{2}\xi_0-1\right)}{3C_\alpha}\, \exp\left(-\sqrt{\frac{2}{3C_\alpha}}\phi\right),\quad r\simeq\frac{64\left( 4H_0^2\xi_0-1\right)^{2}}{3C_\alpha}{\exp}\left(-2\sqrt{\frac{2}{3C_\alpha}}\phi\right).$$
Accordingly with  \eqref{C}  these relations can be presented in the forms:
$$n_s\simeq1-\frac{2}{N_\phi}\exp\left(-\sqrt{\frac{2}{3C_\alpha}}\phi\right),\quad r\simeq\frac{12C_\alpha}{N_\phi^2}{\exp}\left(-2\sqrt{\frac{2}{3C_\alpha}}\phi\right)$$
which are fully correspond to \eqref{prediction}.

\subsection{Restriction to the model parameters}
Accordingly to the Planck data \cite{Akrami:2018odb} value of scalar spectral index and the restriction to tensor-to-scalar ratio are: $n_s\approx0.965\pm0.004$ and $r<0.056$. Value of scalar power spectrum amplitude is $A_s\approx 2 \cdot 10^{-9}$.

The considered inflationary models with the Gauss-Bonnet interaction can be presented more precisely,
namely, to satisfy condition $\epsilon_1(N\simeq0)\approx1$ we should suppose  $N_0=\sqrt{3C_\beta/4}$.
To follow to designations from \cite{Mukhanov:2013tua} we should suppose $N_0=1$ and $C_\beta=4/3$.

Accordingly to \eqref{prediction} the highest value of constant $C_\alpha$ is related with modern observations restriction to the tensor-to-scalar ratio $r$ and the value of e-folding number at the beginning of inflation. At the same time a start point of inflation defines the value of spectral index of scalar perturbations. We numerically estimate the value of the model parameters using \eqref{prediction} and supposing that the inflation begin:
\begin{enumerate}
  \item at $N\approx55-N_0$  before the end of inflation:
$n_s\approx0.964$ and  $0\leq C_\alpha<14.1$;
  \item at $N\approx60-N_0$ before the end of inflation:
$n_s\approx0.967$ and $0\leq C_\alpha<16.7$;
  \item at $N\approx65-N_0$ before the end of inflation:
$n_s\approx0.969$ and  $0\leq C_\alpha<19.6$.
\end{enumerate}
To get  expression for scalar power spectrum amplitude
we substitute \eqref{rG} and \eqref{H2(N)} to \eqref{As}:
\begin{equation}
A_s\simeq {\frac{H_0^{2}{(N+N_0)}^{2}}{6{\pi}^{2}C_\alpha}}\exp\left(-\frac{3C_\beta}{2(N+N_0)}\right)={\frac{H_0^{2}{(N+N_0)}^{2}}{6{\pi}^{2}C_\alpha}}\exp\left(-\frac{2N_0^2}{N+N_0}\right),
\end{equation}
from where
\begin{equation}
\frac{H_0^2}{C_\alpha}=\frac{6\pi^2A_s}{(N+N_0)^2}\exp\left(\frac{2N_0^2}{N+N_0}\right).
\end{equation}
To  estimate ${H_0^2}/{C_\alpha}$  we suppose $N_0\approx1$ in three  cases:
\begin{enumerate}
  \item if the start point of inflation $N\approx54$, then ${H_0^2}/{C_\alpha}\approx 4.09 \cdot 10^{-11}$
  \item if the start point of inflation $N\approx59$, then ${H_0^2}/{C_\alpha}\approx 3.40 \cdot 10^{-11}$
  \item if the start point of inflation $N\approx64$, then ${H_0^2}/{C_\alpha}\approx 2.90 \cdot 10^{-11}$
\end{enumerate}
To estimate relation between model parameters $\xi_0$ and $C_\alpha$ we use relation \eqref{C}:
\begin{equation}
\xi_0=\frac{1}{4}\left(\frac{1}{C_\alpha}-\frac{1}{C_\beta}\right)\left(\frac{H_0^2}{C_\alpha}\right)^{-1}=\frac{1}{4}\left(\frac{1}{C_\alpha}-\frac{3}{4}\right)\left(\frac{H_0^2}{C_\alpha}\right)^{-1}
\end{equation}
in three  cases:
\begin{enumerate}
  \item if the start point of inflation $N\approx54$, then $\xi_0\approx\left(C_\alpha^{-1}-{3}/{4}\right) 6.10\cdot10^9$
  \item if the start point of inflation $N\approx59$, then $\xi_0\approx\left(C_\alpha^{-1}-{3}/{4}\right) 7.35\cdot10^9$
  \item if the start point of inflation $N\approx64$, then $\xi_0\approx\left(C_\alpha^{-1}-{3}/{4}\right) 8.62\cdot10^9$.
\end{enumerate}
The value of parameter $\xi_0$: is positive if $C_\alpha>4/3$; $\xi_0=0$ if  $C_\alpha=4/3$; is negative if $C_\alpha<4/3$.

\section{Conclusion}
We use the equations of the Einstein-Gauss-Bonnet gravity in the Friedmann universe and inflationary parameters
in term of e-folding number for the slow-roll regime. With help of this formulation, we obtain gravity models with the Gauss-Bonnet term  leading to analytical expressions of inflationary parameters coinciding with inflationary parameters of cosmological attractor models in the leading order approximation. The model is a generalization to the cosmological attractor of exponential form initially proposed for general relativity \cite{Mukhanov:2013tua}. We consider the possible expanding of our models for a large field.
We calculate and compare the inflationary parameters for the both models estimate order of accuracies  for large field  expansion.

 Within the framework of the model we obtain an analytical expression for scalar power spectrum amplitude. We estimate relation of constant including to the Gauss-Bonnet term using experimental data for  value of scalar power spectrum amplitude.
 We plan to apply our approach to consideration of more complicated models with the Gauss-Bonnet term and use effective potential proposed in \cite{Pozdeeva:2019agu}.  For future refinement, it should be noted that the presentation of $n_s$ up to second order in $1/N$ can's lead to better agreement with modern observations~\cite{Akrami:2018odb}.\\

 The paper is partly supported by RFBR grant 18-52-45016.

\end{document}